\documentstyle[12pt,aaspp4]{article}

\input psfig

\newcommand{\fun}{\hbox{\ erg cm$^{-2}$ s$^{-1}$} }
\newcommand{\lun}{\hbox{\ erg s$^{-1}$} }
\def \farcs{\hbox{$.\!\!^{\prime\prime}$}}
\def \farcm{\hbox{$.\!\!^{\prime}$}}

\begin{document}

\title{DEEP IMAGING OF AXJ2019+112: THE LUMINOSITY OF A ``DARK CLUSTER''
\footnote{Based on observations collected at the European Southern 
Observatory, La Silla, Chile, proposal number 61.A-0676}}

\author{Narciso Ben\'\i tez\altaffilmark{2}, Tom Broadhurst\altaffilmark{2}, 
Piero Rosati\altaffilmark{3}, Fred Courbin\altaffilmark{4}, 
Gordon Squires\altaffilmark{5}, Chris Lidman\altaffilmark{3},
Pierre Magain\altaffilmark{4,6}} 
\altaffiltext{2}{Astronomy Department, UC Berkeley, 601 Campbell Hall, 
Berkeley, CA}
\altaffiltext{3}{European Southern Observatory, D-85748, Garching, Germany}
\altaffiltext{4}{Institut d'Astrophysique et de G\'eophysique, 
Universit\'e de Li\`ege, 5, Avenue de 
Cointe, B-4000 Li\`ege, Belgium and URA 173 CNRS-DAEC, 
Observatoire de Paris, F-92195 Meudon Principal CEDEX, France}
\altaffiltext{5}{Center for Particle Astrophysics,
UC Berkeley, 301 Le Conte Hall, Berkeley, CA}
\altaffiltext{6}{Also Ma\^{\i}tre de Recherches au FNRS}

\authoremail{benitezn@wibble.berkeley.edu}

\begin{abstract}

  We detect a distant cluster of galaxies centered on the QSO lens and
luminous X-ray source AXJ2019+112, a.k.a.  ~``The Dark Cluster''~
(Hattori et al 1997). Using deep $V, I~$ Keck images and wide-field
$K_s$ imaging from the NTT, a tight red sequence of galaxies is
identified within a radius of $0.2 h ^{-1}$ Mpc of the known $z=1.01$
elliptical lensing galaxy. The sequence, which includes the central
elliptical galaxy, has a slope in good agreement with the model 
predictions of Kodama et al (1998) for $z\sim 1$. We estimate the
integrated rest-frame luminosity of the cluster to be $L_V \geq
3.2\times 10^{11}h^{-2}L_{\sun}$ (after accounting for significant
extinction at the low latitude of this field), an order of
magnitude higher than previous estimates. The central region of the
cluster is deconvolved using the technique of Magain, Courbin \& Sohy
(1998), revealing a thick central arc coincident with an extended
radio source.  All the observed lensing features are readily explained
by differential magnification of a radio loud AGN by a shallow
elliptical potential. The QSO must lie just outside the diamond
caustic, producing two images, and the arc is a highly magnified image
formed from a region close to the center of the host galaxy,
projecting inside the caustic. The mass--to--light ratio within an
aperture of $0.4 h ^{-1}$Mpc is $M_x/L_V=
224^{+112}_{-78}h(M/L_V)_{\sun}$, using the X-ray temperature. The
strong lens model yields a compatible value, $M/L_V=
372^{+94}_{-94}h(M/L_V)_{\sun}$, whereas an independent weak lensing
analysis sets an upper limit of $M/L_V <520 h(M/L_V)_{\sun}$, typical
of massive clusters.

\end{abstract}

\keywords{Gravitational lensing; Galaxy clusters; Dark matter}

\section{Introduction}

  Hattori et al. (1997; hereafter H97) have recently shown that the
well studied QSO lens Q2016+112 is coincident with an extended X-ray
source of hot bremsstrahlung radiation, AXJ2019+112. The association
is based on good positional agreement with high resolution
ROSAT HRI data and also by a plausible consistency in redshift 
between the central lensing galaxy at $z=1.01$ (Schneider et al. 1985)
and an X-ray emission line, interpreted as Fe{\rm XXV} at $z\approx
1$. The detection of enriched gas at high redshift, comparable in iron
abundance to local clusters, is puzzling and exacerbated by the
absence of any obvious cluster of galaxies at this location (Schneider
et al 1985, Langston, Fischer \& Aspin 1991), to provide a potential
means of IGM enrichment.

  Other claims of ``dark''~ cluster sized objects have been made on
the basis of SZ-decrements (Jones et al. 1997, Richards et al. 1997) 
indicating cluster sized plasma concentrations at redshifts in excess 
of unity (Bartlett et al.  1998). However, a follow-up deep X-ray 
observation of PC1643+4631 limits any such plasma to an unreasonably 
high redshift (Kneissl, Sunyaev \& White 1998), and there is no 
evidence for an excess of galaxies toward that direction (\cite{ferr}, 
\cite{saun}). Therefore, confirmation for the 
existence of these high redshift `dark' clusters should await improved 
radio data. 

In contrast, fairly conventional looking X-ray selected clusters have now
been identified at redshifts at high as $z=1.27$ (\cite{stan98b},
Rosati et al. 1998b). These clusters have obvious E/SO color
sequences, identified by optical-IR color and, depending on their
masses, are hard to account for in the standard account of
structure formation with high $\Omega$ (Bahcall \& Fan 1998, Eke et
al. 1998 ). On purely empirical grounds, the existence of X-ray
luminous clusters at $z\simeq 1$ is not surprising. Recently completed
deep X-ray surveys show no evidence of a decline in the space density
of $L_x\sim L_x^*~$ X-ray selected clusters over a wide redshift
range, $0.2<z<0.8$ (Rosati et al. 1998a).

  Here we show that the low galactic latitude of AXJ2019+112 
$(|b|=13.6)$ significantly impedes the detection of faint galaxies in
this field, due to the extended halos of numerous bright stars and
significant Galactic extinction. In \S3 we analyze very
deep, $0\farcs 6$ images taken with the Low Resolution Imaging 
Spectrometer (LRIS) on Keck II (\S2), and identify a
sequence of red galaxies in $V-I$, centered on the lens. This
identification is bolstered by deep $K_s-$band images, obtained using 
the new infrared spectrograph and imaging camera SOFI on the NTT. 
In \S4 we reanalyze the archived ROSAT HRI X-ray observations
of this field, possibly resolving three significant sources of emission, one of
which is marginally extended and coincident with the cluster, leading to a
significant revision of the published X-ray luminosity.  In \S5 we
provide an independent upper limit to the M/L for the cluster from a
weak lensing analysis using the Keck images. In \S6 we deconvolve the
central region of the field, finding an arc around the brightest
cluster elliptical, and finally in \S7 we use this and a simple lens
model to reconcile the many radio sources and optical/IR lensing
features. In \S8 we summarize the main results and conclusions. 

Throughout this paper we assume $\Omega_0=1,
\Lambda=0, h=H_0/(100{\rm~Mpc~km~s}^{-1})$.

\section{Optical and Near-IR Imaging and photometry} 

  Deep $V$ and $I$ images were obtained with LRIS at the KeckII
telescope on the night of 1997 June 29th. The seeing was good,
$0\farcs 62$ in the $I-$band and $0\farcs 65$ in $V$. The images cover
an area of size $5\farcm 5\times 7\farcm 5$.

  The optical images were debiased and flattened with dome flats and
illumination corrected using standard procedures. The $I-$band images
are badly affected by fringing. This was accurately removed by
generating a fringing pattern from combined observations of other
fields taken on the same night. The resulting $V$ and $I-$band images were
aligned with a small correction for a relative distortion across the
field, requiring a third order polynomial. The final, averaged images 
have total exposure times of 1480s in $I$ and 1980s in $V$
with a scale of $0\farcs 215/$pixel and represent the highest
quality optical images of this field to date. Calibration was
obtained in July 1998 by observing the field and several 
Landolt standards with LRIS and the same filter. The magnitude zero point
uncertainties are $0.03$ in both the $I$ and $V-$bands, and no
appreciable color term was detected. Due to the low galactic latitude
of the field ($b=-13.6$), the error in the absolute zero point is
dominated by the uncertainty in the value of the galactic extinction.
\cite{bh} give an extinction of $A(B)=0.98$ at our position,
corresponding to $A(V)\approx 0.74$ and $A(I)\approx 0.44$. The more
direct and higher resolution estimation of \cite{dust}, based on an
IRAS/DIRBE map, gives $A(B)=0.81\pm 0.12$, in marginal
agreement with \cite{bh}.  We adopt the extinction of \cite{dust},
which yields $A(V)=0.67\pm 0.10$, $A(I)=0.36\pm 0.06$ and 
$A(K_S)=0.08\pm 0.01$. Therefore, the total zero point uncertainties are 
$0.1$ in $V$, $0.07$ in $I$ and $0.06$ in $V-I$. The $5\sigma$ limiting 
magnitudes within an aperture of $1\farcs 5$ are $I=25.32$ and $V=26.65$.

In addition wide field $K_s-$band images were acquired on the night of 
1998 July 10th, with SOFI, a new wide field infra--red imager and 
spectrograph on the ESO NTT (Moorwood, Cuby \& Lidman 1998). 
SOFI uses a Rockwell 1024 square Hg:Cd:Te device, which gives a field 
size of $5\arcmin\times5\arcmin~$ in the large field imaging mode. 
The pixel scale is $0\farcs 288$. 

In total, 90 separate images of MG2016 were taken with the $K_s$
filter. Each image is itself a stack of six short
integrations, each of 10 seconds. The telescope
was offset in a random manner by $40\arcsec$ or less between
each stack to help in generating an accurate sky background measurement
over the full field.

 The data was reduced by first subtracting the bias with an averaged
dark frame, and the dividing out the variations in pixel sensitivity
with normalized dome flats. The sky was estimated for each image
individually by constructing a running mean of the adjacent 10
dithered exposures, with suitable clipping to remove objects. The
images were then registered and combined. The intensity of the sky
during the observations ranged between 13 and 13.2 mag/arcsec$^2$.

 The final $K_s$ image is the average of 68 images with a resulting
seeing of $\approx 0\farcs 7$ and a total exposure time of 4080s. This 
image excludes frames with resolution lower than $0\farcs 9$ 
comprising $\approx 25\%$ of the total. The conditions at the time of
the observations were photometric. The magnitude scale was determined
by observing four standards from the list of Persson et al. (1998).
The scatter between the standards is 0.015 magnitudes.  The 5$\sigma$ 
detection within a $1\farcs 5$ aperture is $K_S=21.25$.

\section{Reduction}

  Due to the low galactic latitude of the field, the star density is
very high, so that most of the area of the field is covered by the
extended haloes of the brightest stars, especially in the $I-$band 
(Fig \ref{fig1}a). 
We find that the best approach to dealing with this problem is to
generate a ``halo''~ pattern by setting the central parts of the stars
and the saturation spikes to the average background level and then
applying a broad, $11\arcsec\times 11\arcsec$, median filter to the
image.  The pattern thus obtained is subtracted from the original
frame and the resulting image used for object detection. In a final
step, all the bright cores of the stars, including their saturated
trails, are masked out.  The initial and final, `de-haloed' $I-$band
images are shown for comparison in Fig 1a,b.

  The smoothing and masking affects the photometry of bright
extended objects, but on average it does not affect the fainter ones.
This is established using similarly deep but much higher latitude
Keck $I-$band observations of a control image PSR 1640+22 field
(\cite{psr}). This field was scaled so that the background noise level
was the same as in our Q2016+112 $I$ image. Then we add it to all the
haloes and bright objects from our cluster field, i.e. the difference
between Fig 1a and Fig 1b. In this way, a test image is created with
the same `halo' characteristics as Fig 1a but with a known
`uncontaminated' galaxy background for performing tests.  We then
apply exactly the same halo removing procedure described above. In
establishing the errors introduced by the haloes and the improvement
from the halo removal procedure we make use of the detection software
SExtractor (Bertin \& Arnouts 1996) to produce three
catalogs. Firstly we create a catalog of the the photometric
properties of the test field before any changes are made (catalog O). 
Secondly a catalog is created for the halo contaminated control (catalog H)
field by adding the halos from the cluster frame to the control field, and 
finally a catalog (dH) is generated for the halo subtracted control field after
applying our correction for the haloes in exactly the same way as for the
cluster field.

\begin{figure}[t]
\caption{$I-$band image of the field around Q2016+112 before (a) 
and after (b) the halo removing procedure. The white circle shows 
a $0.4h^{-1}$ Mpc diameter aperture ($47\arcsec$) at $z=1$ centered on 
galaxy $D$}
\label{fig1}
\end{figure}

  Comparing the test field catalog (O) with those of the halo affected
images of this same field (H and dH) allows a reliable empirical estimate 
of the effect of the haloes on the magnitude estimates and also measures 
the degree of improvement achieved by our halo removal
procedure. The comparison of the magnitude differences 
between catalogs (H) and (O), and catalog (dH) and (O) shows that the
presence of numerous stars with extended haloes distorts the
estimation of the background, and is a stronger source of photometric
noise than any additional uncertainty in eliminating the
haloes. These tests also allow us to estimate the percentage of
`useful surface' recovered with our procedure. On the image obtained
after adding the haloes we only detect between $35\%($at $I\approx
20)$ and $25\%$ at $(I\approx 25.5)$ of the objects which were
initially present on the PSR 1640+22 image. After the halo removing
procedure, we recover respectively between $30\%(I\approx 20)$ and 
$45\% (I\approx 25.5)$ of the same objects.  The maximum possible 
detection efficiency (found by excluding from the PSR 1640+22 original 
frame all objects which have at least one pixel in the masked areas) 
is $44\%($at $I\approx 20)$ and $72\% (I\approx 25.5)$, in the last case
approaching the percentage of `non-masked' area, $74.5\%$. This means
that apart from obtaining slightly more reliable magnitudes, the
followed procedure increases by $\approx 80\%$ the useful surface for
detecting faint objects.  Fortunately, as can be seen in Fig 1,
the central region containing the cluster is one of the least affected
by haloes. But in the rest of the image, in particular areas close to
bright haloes, the tests show that apart from the standard photometric
errors there is an increased random error in the magnitude estimates
of $\approx 0.2$ in the $I-$band magnitude compared to the
`halo-free' image.

  The detection and classification of objects in the $V,I$ and $K$ images
is made using SExtractor v2.0 (Bertin \& Arnouts 1996). This software
smooths the images with a gaussian of size close to the seeing width,
and then locates peaks above a given threshold. To obtain accurate
colors we use the $I-$band image for detection and use the apertures
obtained for each of image to perform the $V$ and $K_S-$band
photometry. These matched apertures are ideal given the very similar
seeing for all 3 passbands.
 
  It was found that somewhat different definitions of magnitudes are
required for bright and faint galaxies for best results, due to the
approach of all faint galaxies to the seeing profile. For the bright
galaxies we adopt the SExtractor Kron-like `automatic' magnitudes
$m_{au}$, with a $k$ parameter of 2.5.  These apertures encompass
$\approx 95\%$ of the total object magnitudes (Bertin 1998). To prevent the
magnitudes from being affected by crowding, SExtractor masks the
pixels belonging to neighboring objects and replaces their values by 
those of pixels lying symmetric opposed to the source center. For faint
objects ($I>23.5$), we measure the magnitude within a fixed aperture
of $1\farcs 5$ and correct it to the automatic magnitude assuming
that faint objects have roughly stellar profiles. This was established
by stacking together many faint objects. The corresponding corrections
are $0.14$ in the $I-$band and $0.21$ in the $V$ band.
	 
\begin{figure}[t]
\epsscale{0.65}
\plotone{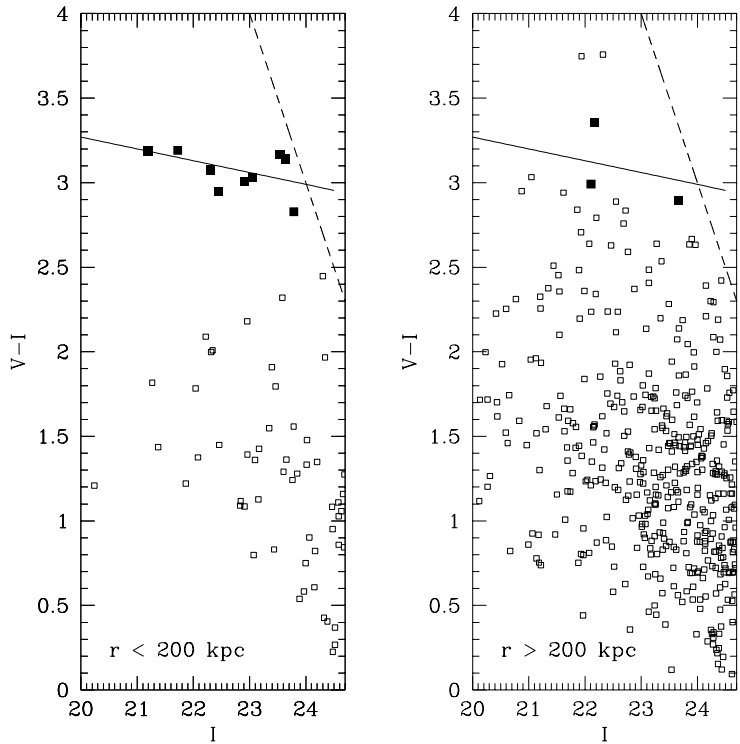}
\caption{Color magnitude diagram for the central $0.4h^{-1}$Mpc 
region around Q2016+112 (left) and for the rest of the frame. 
The filled squares indicate the positions of the `red sequence' galaxies;
objects compatible with having 
$V-I=3.19-0.07\times(I-21.20)$ within errors} 
\label{cmdia}
\end{figure}

  SExtractor assigns to each object a value of the parameter
``class-star'', which ranges from 0 to 1. If the seeing FWHM is well
determined, stars obtain values of this parameter close to 1, whereas
for extended objects it tends to be close to 0. The classification,
performed using the $I-$band images, works satisfactorily for objects
with $I\lesssim 24.7$, which corresponds to a detection with a
$S/N\gtrsim 9$ within a $1\farcs 5$ aperture. At fainter magnitudes,
SExtractor assigns similar values of the ``class-star''~ parameter
$\approx 0.4-0.5$ to all objects. One of our main concerns is to avoid
the contamination of our galaxy sample by stars, specially some of them, 
e.g. disk M dwarfs, would have the same colors as early-type
galaxies at the cluster redshift. This is achieved by excluding those
objects with $I>24.7$ and imposing a very conservative threshold in
the ``class-star''~ parameter, $0.25$.  The $V-I$ vs. $I$ plots of the
central region around the lens and the rest of the frame are shown in
Fig \ref{cmdia}. Objects with $V$ fainter that $V=27.0$, corresponding
to a $3.5\sigma$ detection within a $1\farcs 5$ aperture, have been
excluded. Fig \ref{vik} shows the $V-I$ vs $I-K$ for all the galaxies
in the field. Objects fainter that $K_S=21.60$.  corresponding to a
$3.5\sigma$ detection within a $1\farcs 5$ aperture, have been excluded.

\section{The galaxy cluster}

  Recent work (\cite{stan98a}, \cite{koal}) has shown that cluster `red
sequences', the locus in the color--magnitude diagrams of the cluster
early type galaxies, represent a remarkably stable feature which is
detected up to $z>1$ redshifts (e.g. on the $z=1.2$ cluster around
3C324, Dickinson 1995). The red sequence is an
excellent marker of the presence of a massive cluster, since galaxies
with the colors and magnitudes of cluster early types are scarce among
the general field population. 

\begin{figure}[t]
\plotone{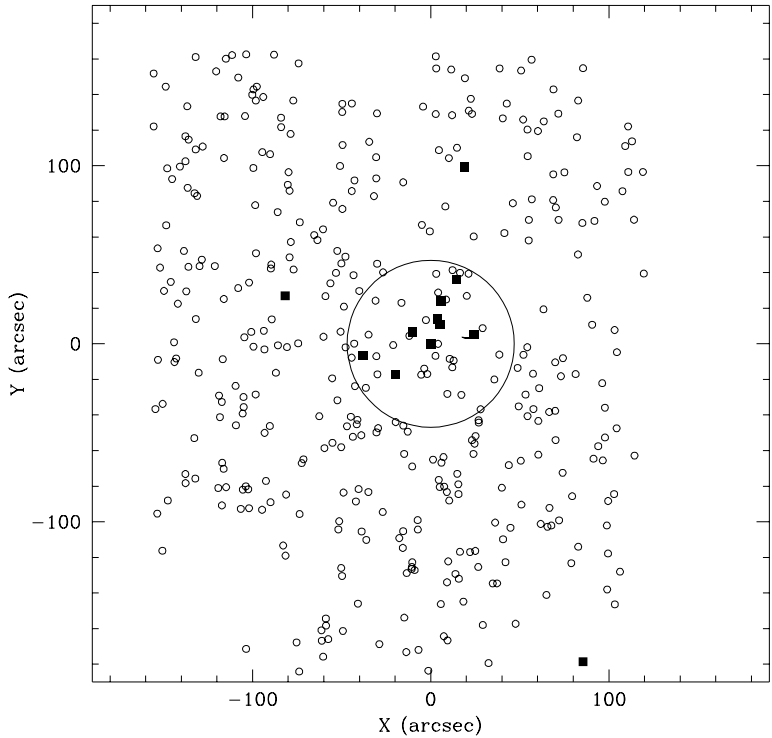}
\epsscale{0.65}
\caption{Spatial distribution of the cluster early type candidates, 
represented by filled squares. The empty squares represent the stars 
detected on the frame. The circle corresponds to a $0.2h^{-1}$ Mpc 
radius at z=1}
\label{spdis}
\end{figure}

\begin{figure}[t]
\epsscale{0.65}
\plotone{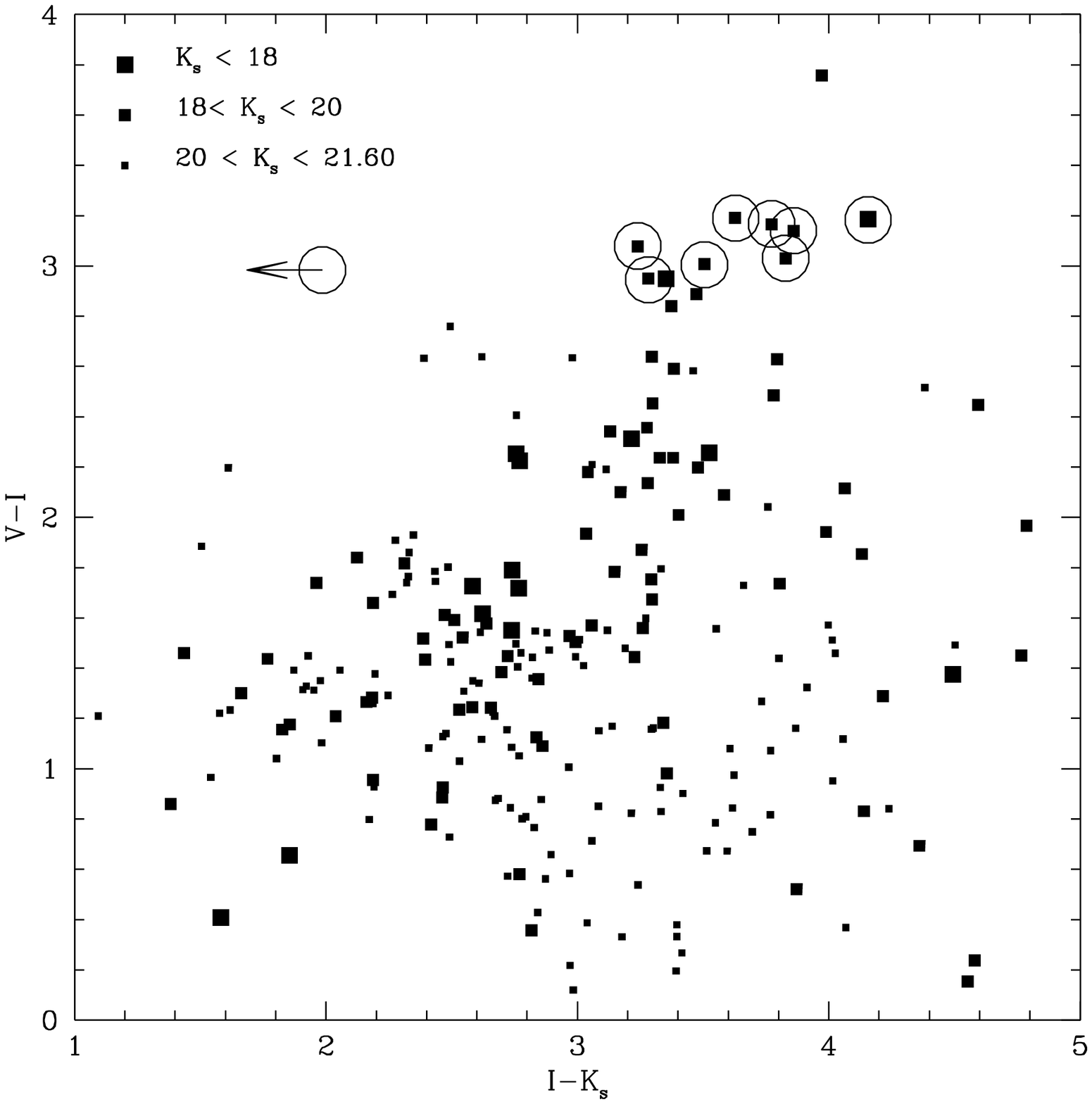}
\caption{$V-I$ vs $I-K$ plot for all the galaxies in the field. 
The objects identified as cluster members on the basis of the 
$V-I$, $I$ plot are circled. The arrow indicates an upper limit }
\label{vik}
\end{figure}

 To examine the data for evidence of a cluster sequence, we have
divided the frame into two regions, one corresponding to a
0.4h$^{-1}$Mpc circular aperture at $z=1.01$, centered on the luminous
galaxy $D$ (Fig \ref{last}b), and an 
outer region covering the rest of the image. Using
the magnitude of galaxy $D$ as its zero-point ($I=21.20$, measured using
the deconvolved images obtained in Sec 6), and adopting a slope of
$0.07$ at $z=1.01$ (\cite{koal}), the early-type sequence of such a
cluster should lie at $V-I=3.19-0.07(I-21.20)$. Fig \ref{spdis} shows
the positions of the galaxies with magnitudes compatible within their
observational error with this relationship, and those of all the stars
in the frame (objects with class-star higher than 0.95), independently
of their color. It is obvious from this figure that a red
sequence of galaxies is clustered around $D$ and also that galaxy $D$
lies on this sequence. The central region has 
$8$ such candidates (excluding galaxy $D$), whereas there are only
other $3$ such galaxies on the rest of the frame. The central circle
contains $54$ stars against $390$ detected in all the frame, i.e. the
central region comprises only 
$13.8\%$ of the `useful' surface in the frame. Thus, there is a factor
$\approx 20$ overdensity of red galaxies in the central $0.4$ Mpc, 
clear evidence of the existence of a galaxy cluster. A 
rough estimation of the significance of this result may be
performed using a binomial distribution. The probability of having $k$
galaxies out of a total of $n$ within a region containing a fraction
$p$ of the total surface is represented by
$p(n,k)=C_n^kp^k(1-p)^{n-k}$. With $n=11$, $k=8$ and $p=0.138$, we
find $p(11,8)=1.4\times 10^{-5}$, corresponding to a confidence level 
of $>99.998\%$.  Fig \ref{vik} shows that all but one of the cluster
member candidates selected on the basis of the $V-I$, $I$ are also 
very red in $I-K_s$. This further supports
that these galaxies are physically associated. Independently, in 
addition to galaxy $D$, two of our sequence galaxies have spectroscopic
redshifts, and both are claimed to lie at $z=1.0$ (Kneib et al. 1998).

In the remainder of this paper, the galaxies listed in table 1 
(except the one which is not detected in the $K-$band) will be 
considered cluster members. To estimate their 
absolute magnitudes $M_V$, we use a k-correction of $3.5$ 
at $z=1.01$ for evolved ellipticals 
(Poggianti 1997), and a distance modulus of $42.75$.
Galaxy $D$ has a de-reddened luminosity of 
$L_V^D=4.16\pm 0.21\times 10^{10}h^{-2}L_{\sun}$, 
and the total luminosity of the detected cluster candidates is 
$L_V=1.33\pm 0.08 \times 10^{11} h^{-2}L_{\sun}$. 
Our detection limit of $V=27$ corresponds to 
$M_V=-19.25$, so to compare this luminosity with the results of \cite{mass} 
(hereafter S97), we correct it by integrating a Schechter function up to 
$M_V=-18.5$, using $\alpha=-1.25$ and $M_V^*=-20.75+5\log h$, measured by 
S97 at $<z=0.55>$. The correction amounts to $20\%$ of the total 
light, and it is quite robust with respect to the exact value of 
$M_V^*$, ranging from $30\%$ for $M_V^*=-20.30$ to $10\%$ for 
$M_V^*=-21.60$. The adopted total luminosity for 
the cluster early type population is thus $L_V^E=1.6\pm 0.2\times 
10^{11}h^{-2}L_{\sun}$. In the cluster sample of S97
the early types contain typically $50\%$ of the total cluster light. 
Therefore, the total luminosity of the `dark' cluster would be 
$L_V^{\rm all} \approx 3.2\times 10^{11}h^{-2}L_{\sun}$. This value 
probably underestimates the cluster luminosity, as the fraction 
of cluster galaxies with star formation, and therefore bluer than the 
red cluster sequence, is likely to be higher at $z=1$ than at 
$z\lesssim 0.5$.

 H97 quote a value of $L_B=2.75\times 10^{10}h^{-2}L_B^{\sun}$ for
galaxy $D$. As no other obvious cluster galaxy candidates were known,
they adopted that value as the total luminosity. Assuming a color of
$B_J-V\approx 0.6$ for an elliptical at redshift $z=1$, their value of
$L_B$ would correspond to $L_V\approx 3 \times 10^{10}h^{-1}L_{\sun}$,
an order of magnitude lower than our estimate of the total cluster light.

\section{X-ray analysis}

\begin{figure}[t]
\plotone{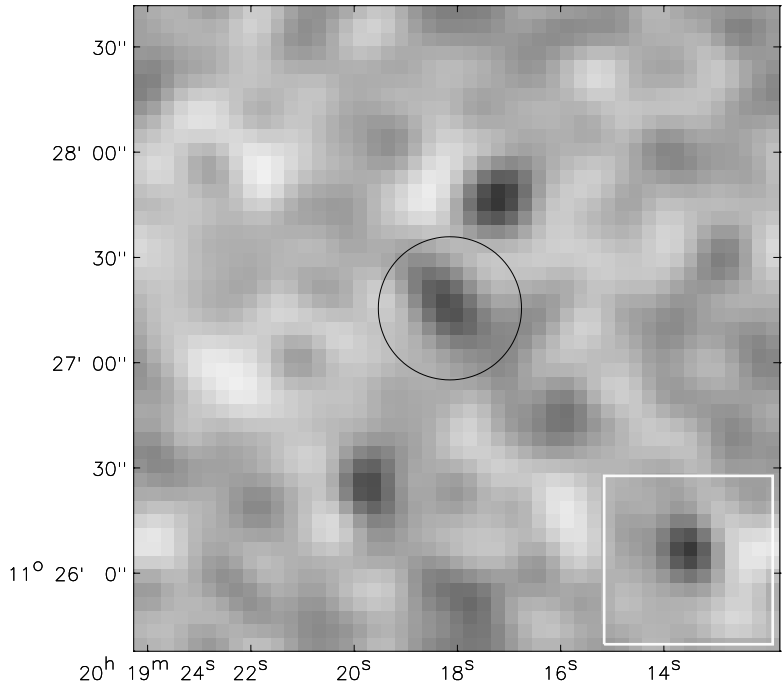}
\epsscale{0.65}
\caption{Central region of the ROSAT-HRI field smoothed with a gaussian
filter with $\sigma = 6\arcsec$ to match the resolution of the HRI data. 
The source in the lower right box is a
simulated point-like source with the same flux encircled by the central
source, to show the difficulty of resolving the central X-ray emission.}
\label{xray1}
\end{figure}

\begin{figure}[t]
\plotone{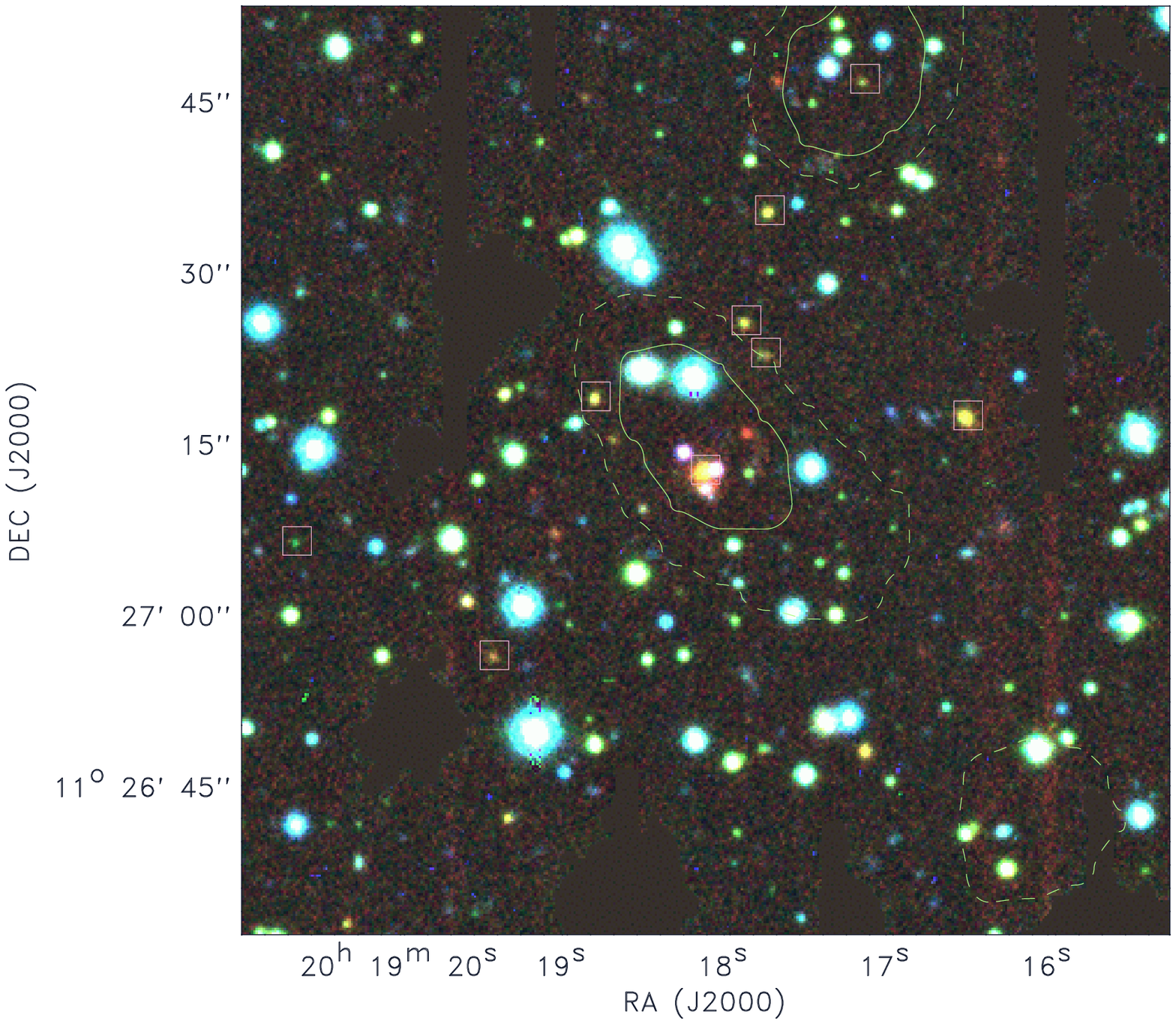}
\epsscale{0.65}
\caption{ROSAT-HRI contours overlaid on the optical-IR $(VIK)$ image
of the MG2016 field. The contours are at 2 (dashed) and 3 $\sigma$
levels (smoothing gaussian scale  $\sigma = 8\arcsec$). The boxed
objects the cluster galaxies candidates belonging to the red $V-I$ 
sequence listed in Table 1. This figure demonstrates that the 
red galaxy overdensity is spatially coincident with the central X-ray 
emission. 
}
\label{xray2}
\end{figure}

  We have reanalyzed the ROSAT-HRI archival observations to examine
the spatial relationship between the optical and X-ray emission. The
HRI dataset consists of three separate pointings with ROSAT ID:
rh800811n00, rh800811a01, rh800811a02 with a total exposure time of 79
ksec. The X-ray photon files have been processed with standard
IRAF-PROS routines filtering out time intervals with high background
rate. The resulting coadded image with $4\arcsec$ pixel size has an
effective exposure time of 56.0 ksec, mostly contributed from the A01
and A02 observations. The astrometrical accuracy is found to be better
than $5\arcsec$ using 2 bright X-ray sources coincident with bright
stars.  A faint source at $4\sigma$ level is detected within
$3\arcsec$ of the position of the elliptical $D$. A second source
($\sim\!  4.5\sigma$) is found $30\arcsec$ NE.  The latter seems to be
missing in the original analysis by H97. However, inspection of
several images obtained by splitting the observation in two time
intervals reveal variations in the flux of this source, suggesting
that it might be either variable or spurious, whereas the central
source shows a consistent flux.  Furthermore, the elliptical
morphology of the central X-ray source is present in two independent
pontings, suggesting that it is not an artifact of the background
noise. This central source is the natural X-ray counterpart of the
optical cluster and is likely coincident with the ASCA source of
H97. The central 3 arcmin of the HRI field are shown in Fig
\ref{xray1}.  Corresponding X-ray contours are overlaid on the $VIK$
color image of the field in Fig \ref{xray2}.

\begin{figure}[t]
\plotone{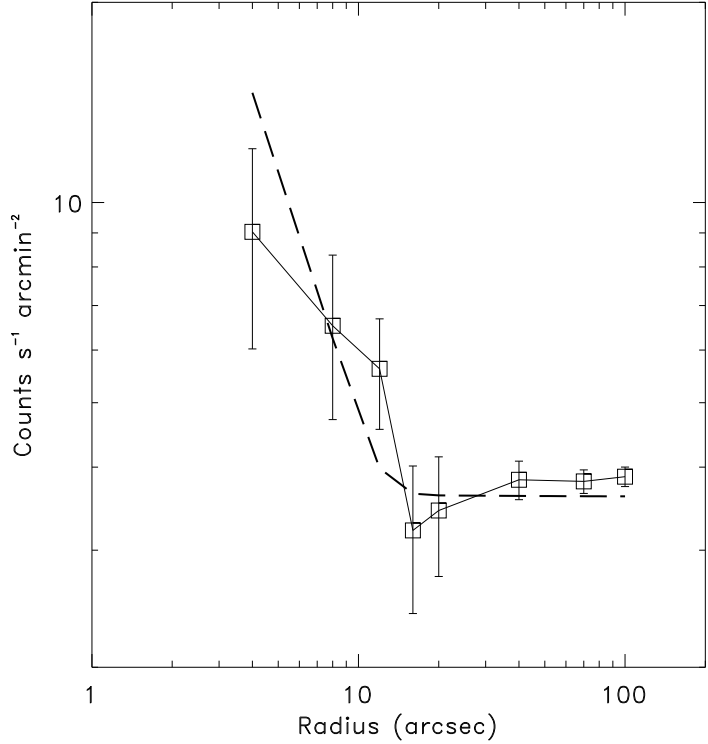}
\epsscale{0.65}
\caption{X-ray surface brigthness profile of the central HRI source. The HRI
 noiseless PSF for a source of the same flux is superimposed (dashed
 line). The background is equal to $\sim\! 3.6$ counts s$^{-1}$ arcmin$^{-2}$.
 The possibility that this emission is unresolved is hard to exclude. }
\label{xray3}
\end{figure}

  We find $17\pm 9$ net counts, or a count rate of $(3.0\pm 1.6)\times
10^{-4}$ cts s$^{-1}$ in the central source within a circle of
$20\arcsec$ radius ($\simeq 85$ h$^{-1}$~kpc at $z=1$). The flux
within a circle of $1\arcmin$ falls short of the H97 estimate by a
factor of two. Using a galactic hydrogen column density of
$N_H=1.5\times 10^{21}$ cm$^{-2}$ and a Raymond-Smith spectrum with
$T\simeq 8$ keV, as measured by H97, the program PIMMS for the HRI
yields that 1 cts/s is equivalent to an unabsorbed flux of $6.06\times
10^{-11}\fun$ in the $0.1-2.0$ keV band. The flux of the
central source is $F_x(<20\arcsec) = 1.8 \times 10^{-14}\fun$, or an
X-ray luminosity at $z=1$, $L_x(<20\arcsec) = 1.1 \times 10^{44}\lun$
in the observed 0.1-2 keV band.  The K-correction amounts to a
factor 1.4, whereas the conversion factor for the 2-10 keV band is
$\sim\! 1.6$. Therefore, $L_s(<20\arcsec) = (1.1/1.4\times 1.6)=1.3
\times 10^{44}\lun$ in the 2-10 keV rest frame band. This value is a
factor 3-4 lower than the estimate by H97, even with extrapolation of
the flux out to $1\arcmin$ using their assumed surface brightness
profile.  Given the low signal-to-noise of the central source, it is
not possible to assess its angular extent. This is shown visually in
Fig \ref{xray1}, where the HRI PSF has been used to simulate a source
of 17 counts and added to the central region. The comparison between
the surface brightness profiles of the central object and the
point-like (noise-less) source is shown in Fig \ref{xray3}, although
at these low S/N levels, profile fitting is rather meaningless. The
spatial analysis of H97 suggested an extended central source. This
might be due to the large smoothing scale (FWHM$=35\arcsec$) used by
those authors or to a higher background of our combined X-ray image.

 The fact that the X-ray emission of a cluster at $z=1$ is unresolved
with the HRI is not surprising, given the severe surface surface
brightness dimming at this redshift (Rosati et al. 1998a).  However, we
cannot rule out a significant contribution to the X-ray emission
from the two lensed QSO images. Therefore, the aforementioned X-ray
luminosity should be considered an upper limit. Improved observations,
ideally spatially resolved spectroscopy, should help to clarify the
relation between the hot thermal spectrum observed by ASCA, with is
low spatial resolution, and the patchy nature of the much better resolved
ROSAT-HRI map.

 Assuming that the cluster is isothermal and in hydrostatic
equilibrium, Hattori et. al. estimate the gravitational mass
within a radius of $0.5h^{-1}$ Mpc to be $M=1.8^{+0.9}_{-0.6}\times
10^{14}h^{-1}M_{\sun}$, where the errors represent the uncertainty in
the cluster temperature determination $kT=8.6^{+4.2}_{-3.0}~$keV. 
Assuming $M \propto r$, the mass inside a
$0.2h^{-1}$Mpc radius would be $M_x=7.2^{+3.5}_{-2.5}\times
10^{13}h^{-1}M_{\sun}$. This gives a mass to light ratio for AXJ2019+112 
of $M_x/L_V^E=448^{+224}_{-153}h(M/L_V)_{\sun} $ and 
$M/L_V^{all}=224^{+112}_{-78}h(M/L_V)_{\sun}$, compatible with 
the median values of $z\lesssim 0.5$ clusters (S97): 
$M/L_V^E=330^{+210}_{-110}h(M/L_V)_{\sun} $ and 
$M/L_V^{all}=180^{+140}_{-80}h(M/L_V)_{\sun} $.

\section{Weak lensing Analysis}

  The observations are of sufficient depth and resolution to allow a
useful weak lensing analysis. The procedure for measuring the lensing signal
is standard, and follows that employed in other cluster lensing
studies (e.g., \cite{squires96}). Object finding and analysis employed
the IMCAT software developed by N. Kaiser (\cite{ksb95} -- KSB).
Correction for psf anisotropy, and circularization due to seeing was
made with the standard formalism of KSB and Luppino/Kaiser (1997),
with corrections noted in Hoekstra et al. (1998). Correction for these
systematics is very robust, due to the very regular psf in the final
coadded images, and the overabundance of stars in this field.
The residual ellipticity in the corrected star catalog is less than
0.1\% over the full field and so residual systematic shape distortion
in the galaxy catalogs is completely negligible.

  Using the standard unbiased inversion algorithm (e.g., \cite{ks94};
\cite{sk96}), and the $\zeta$-statistic, we determine the cluster
surface density from the corrected galaxy shape measurements. For $I
< 25$ we find a {\em null} detection of the cluster mass
distribution, both in a 2D mass map, and in the radially averaged mass
profile.

  This null detection allows us to place an interesting upper limit on the
cluster mass, albeit in a model dependent fashion.  For example,
assuming the cluster follows a isothermal profile, $\rho(r) = \sigma_v^2
/ 2 \pi G r^2$, where $\sigma_v$ is the line of sight velocity
dispersion, our data are of sufficient depth to detect the signal from a
$1000$ km/s singular isothermal sphere (SIS) cluster at the 
$3\sigma$ level at $r = 1'$ and a
$2\sigma$ level at $r = 2'$, for $\beta=0.2.$ In terms of mass, for a
cluster with $\sigma=1300~{\rm km \; s}^{-1}$ (typical of clusters with
$T_x\approx 8.5$~keV) this model yields
\begin{equation}
M_{3D} < 1.6 \times 10^{14} \left(\frac{\sigma}{1300 \;
{\rm km \; s^{-1}}}\right)^{2} \left(\frac{\beta}{0.2}\right)^{-1}
\left(\frac{r}{ 200 \; h^{-1}\;{\rm kpc}}\right) h^{-1} M_{\sun}
\end{equation}
where $\beta$ describes the average relative distance between the lens
and the source galaxies. For $I<25$ and for plausible cosmologies and
star formation models, $\beta$, ranges between $0.16$ and $0.23$, so the
above result should be quite robust.  This result sets the constraint 
$M/L_V^{all}<520h(M/L_V)^{\sun}$ independently of the X-ray and 
strong lensing estimates.

\section{Image deconvolution}

\begin{figure}[t]
\plotone{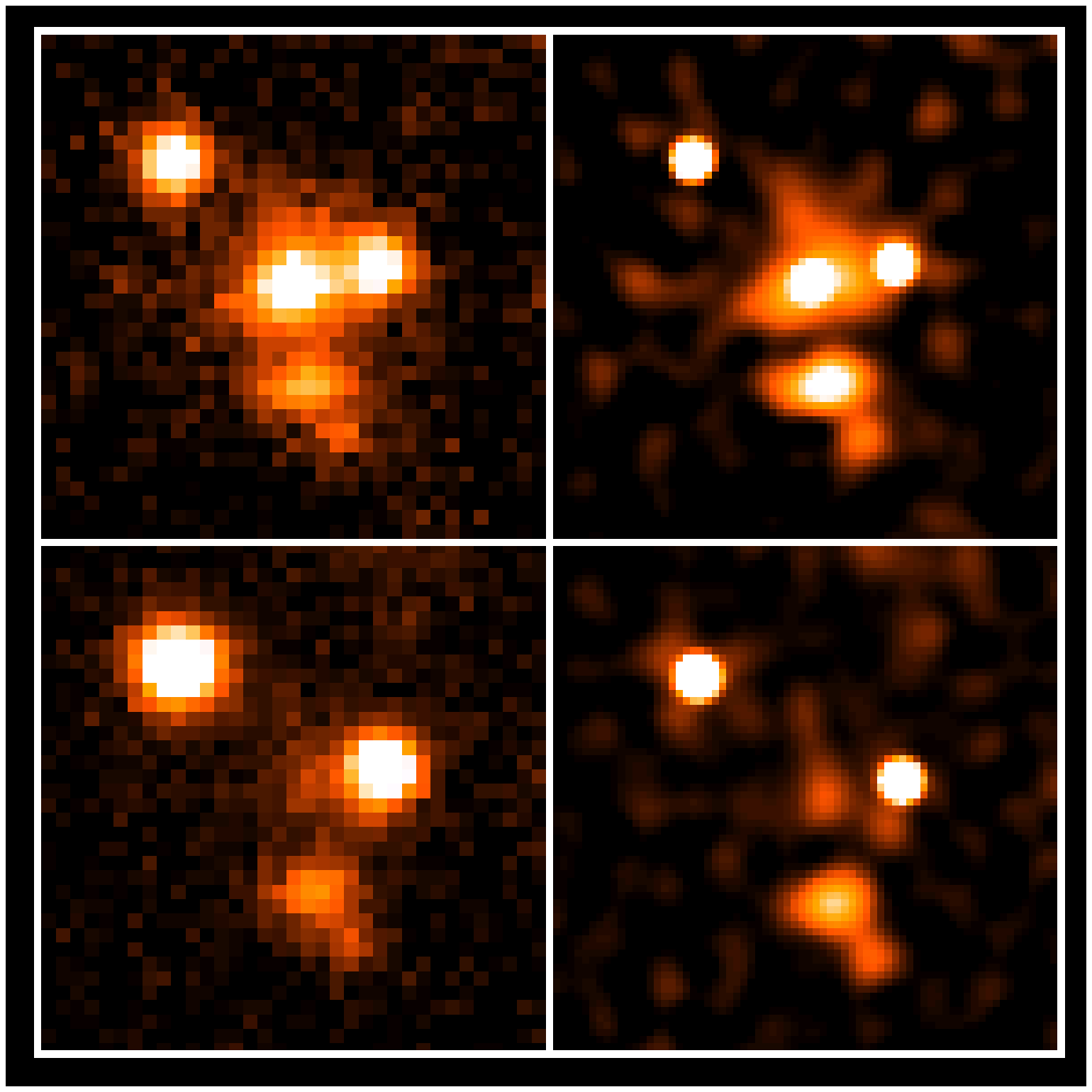}
\epsscale{0.3}
\caption{(Top) Stack of four $I$ band Keck/LRIS images of Q2016+112
with  a seeing of $0\farcs 6$  and  pixel  size of $0\farcs 215$.  The
simultaneous  deconvolution of the four  frames is  shown on the right
with a  resolution improved to  $0\farcs 32$  and a finer pixel size of
$0\farcs 108$. (Bottom) Stack of four $V$  band images and  their
simultaneous deconvolution. Note the obvious arc feature lying below the
two QSO images which modeling shows is an image of the central
region of the host galaxy of the QSO. This arc is resolved in both 
the radial and tangential directions, indicating a flat central potential.}
\label{fred}
\end{figure}

  The data can also be usefully used to study the immediate
environment of the lens, which shows many interesting lensing
features.  First we deconvolve the Keck $V$ and $I$ images using the
MCS algorithm (Magain, Courbin \& Sohy  1998).  Since several
individual dithered frames were available, the simultaneous
deconvolution capability of the algorithm can be applied to improve
not only the resolution of the images, but also the sampling.  A total
of four images were used in the $V$ and $I$ bands, which allows an
effective reduction in the pixel size of the data to $0\farcs 108$ in
the deconvolved combined image compared with the intrinsic
instrumental scale of $0\farcs 215$ per pixel.  In principle, the
fundamental limitation in resolution is imposed by the Nyquist
frequency set by the {\it adopted} sampling in the deconvolved frame,
i.e.  2 pixels FWHM. Since the present images show sharp, high
frequency structures, we found it reasonable to limit ourselves to a
resolution of 3 pixels FWHM or $0\farcs 32$ in both bands and avoid to
recover spurious high frequency spatial structure too close to the
Nyquist limit. Note that this strict adherence of the deconvolution to
this fundamental limitation distinguishes the MCS method from other
deconvolution procedures where sharp noisy backgrounds are formed,
generating spurious structure.

  The simultaneous deconvolution of the frames was performed in the
same way as in Courbin et al. (1998a,b).  The final deconvolved frames
are displayed in Fig. 8 together with the stacks of the 4 original
images. Given the low galactic latitude of Q2016+112, many stars are
available in the immediate vicinity of our target.  Accurate PSFs can
therefore be constructed for all the images in the stack. The residual
maps obtained after deconvolution were satisfactory for all frames in
the $V$ and $I$ datasets.  However, because the central part of the
main lensing galaxy (object $D$) is very sharp in the $I$ band, we
choose to model it as a point source plus a diffuse background.  This
certainly does not mean that the core of the lens is actually a point
source, we model it as such to exclude the spurious high spatial
frequencies it would otherwise introduce.

  The deconvolution procedure returns the astrometry and photometry of
the point sources and decomposes the final image into a sum of point
sources plus a diffuse background. A major advantage over other
deconvolution algorithms is that the resulting deconvolved images
allow accurate aperture photometry for the extended objects,
free of significant light contamination by bright point sources such
as QSOs. This decomposition capability allows us to derive accurate 
magnitudes for the lensing galaxy, especially in the $V$ band since it 
is heavily contaminated by image B at this wavelength. 

 A summary of the photometry is given in Table 2, including the well
resolved bright arc apparent below the two QSO images as discussed
further below.

  Recent NICMOS2 archival images (ID: 7495, P.I: Falco, E.) are
consistent with the above deconvolution in the near-IR, showing very
clearly the main arc below the central luminous galaxy, despite the
poor point spread function of images taken prior to the refocusing of the instrument,
indicating the arc is relatively red.  The H band (F160W ) image is
shown in Fig \ref{last}a for comparison. This extended IR feature was
also noted by \cite{lan91} in high quality ground based IR images.

\section{Lens Model}

\begin{figure}[t]
\plottwo{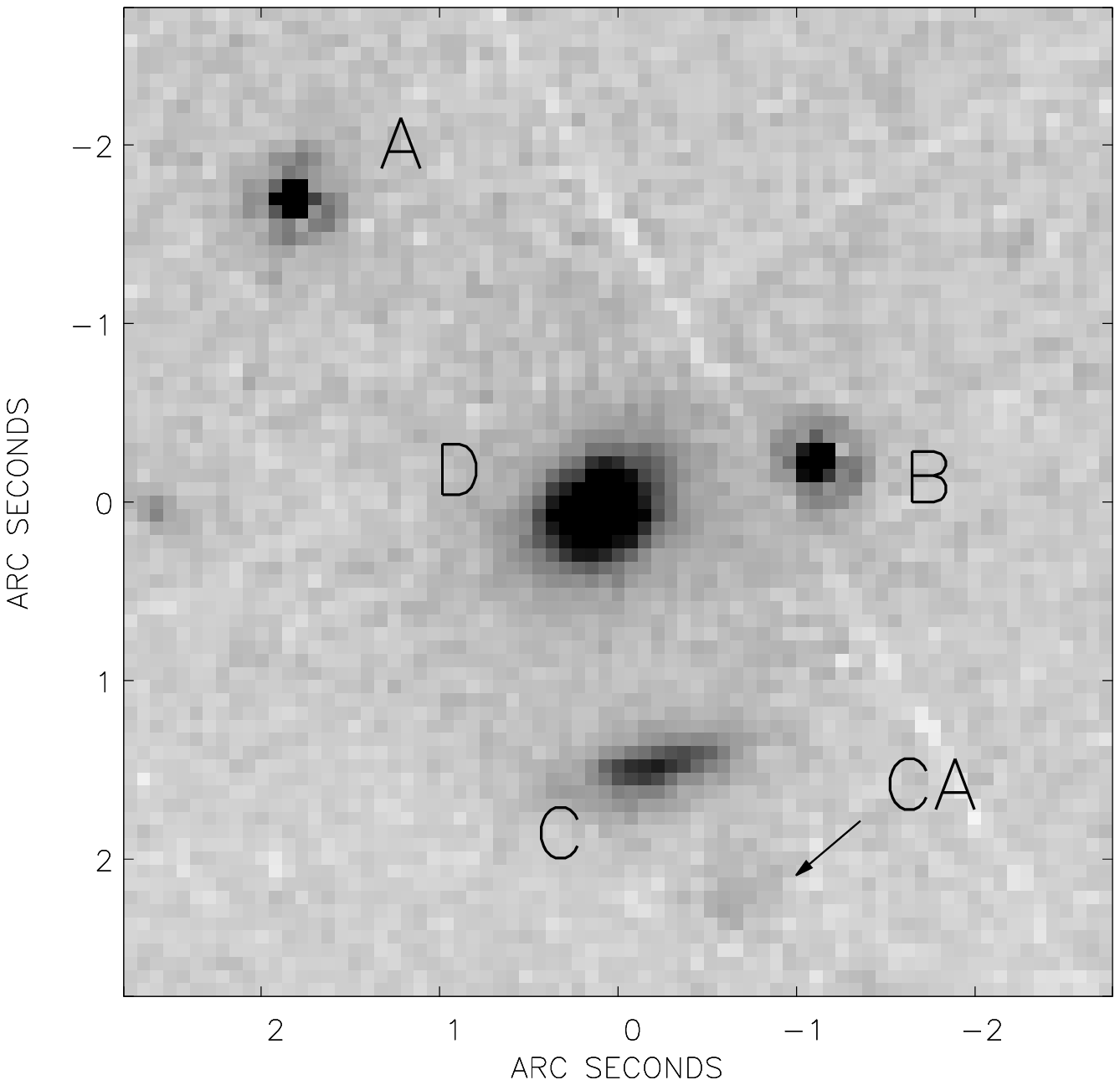}{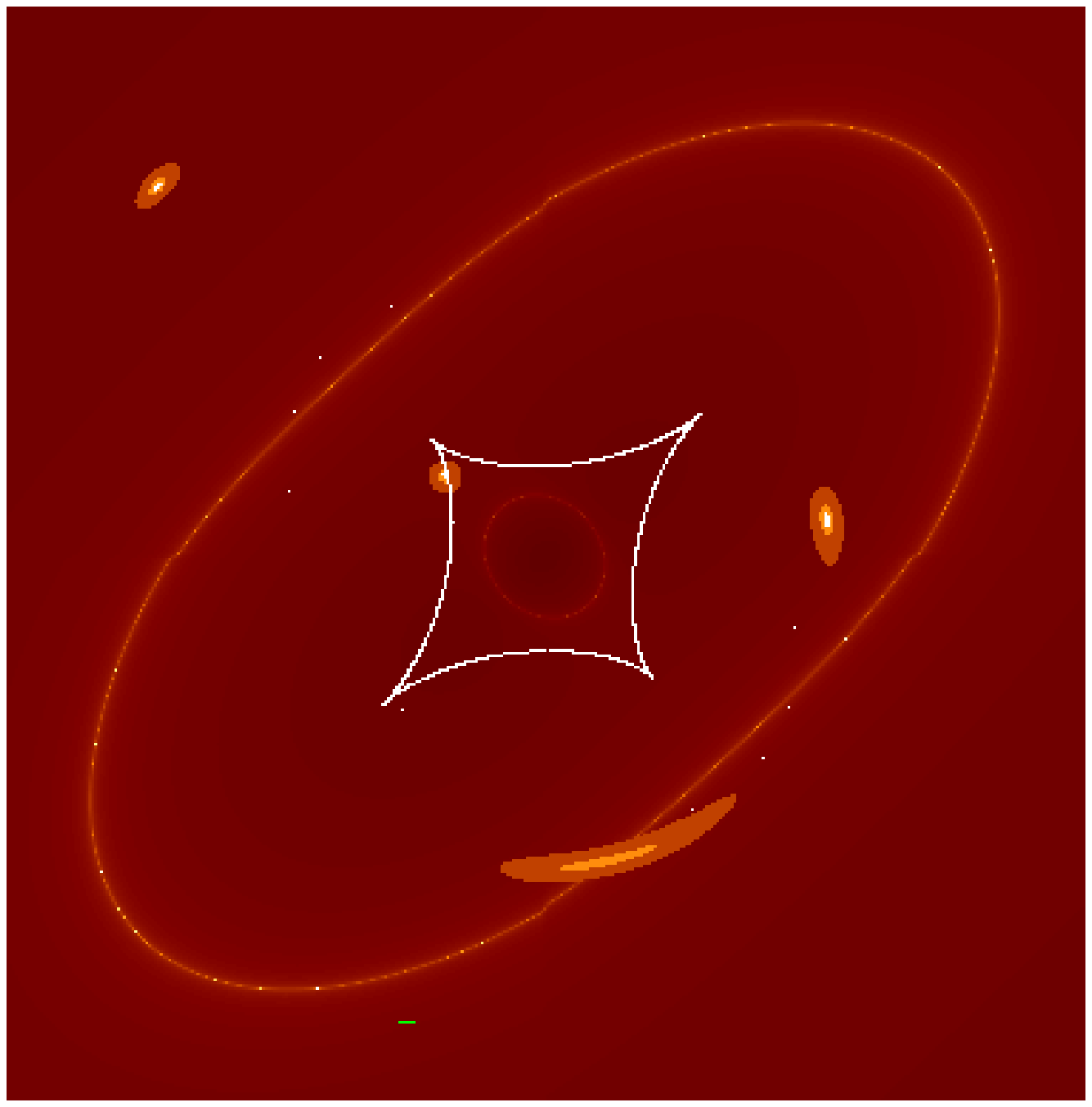}
\caption{The left figure corresponds to the NICMOS H band image of 
MG 2016+112. To the right, the lens model with the source, critical 
curve and caustic superposed on the magnification field. The colors of 
the source represent three regions; the bright central spot indicates 
the location of the QSO which must lie just outside the caustic generating 
only two images, and the lower surface brightness region 
crossing the caustic represents the central host galaxy which 
forms the optical-IR arc. The center of the arc is coincident with the  
extended radio source, C, this radio emission maps back to a 
region of the source immediately inside the caustic, 
very close but distinct from the central QSO.}
\label{last}
\end{figure}

  The lens was originally discovered in the radio, as a close pair of
bright flat spectrum sources (Lawrence et al 1984) separated by
$3\arcsec$.  These bright radio components, $A$ and $B$, are accurately
coincident with optical QSOs (Schneider et al. 1985).  A third radio
image, $C$, (Fig \ref{last}a) lies nearby and is resolved with VLBI into 
4 knots along a line of only 0.1 arcsec. These knots have a flatter 
spectrum than $A$ or $B$, so that their relation to the QSO is not 
clear (\cite{gar94}, \cite{hef91}, Nair \& Garrett 1998). 
In addition, image $C$ is coincident with the center of
an arc found in $K-$band imaging (\cite{lan91}) and clearly
confirmed in an HST NICMOS exposure shown in Fig \ref{last}a . The
optical counterpart of their $K-$band arc is clearly seen in our
deconvolved images, which show that it is resolved in the radial as
well as the tangential direction. The redshift of both the lens and the
source were measured by Schneider et al. (1985) to be $z=1.01$ and
$z=3.2$ respectively.
 
  Here we show that the various images are relatively simply related
if the lensed QSO and its host galaxy lie close to the caustic of the
lens, so that different parts of the source are magnified by different
amounts.  The lens model is kept as simple as possible. In addition to
the usual choice of variables, i.e. position, position-angle,
surface-density normalization and mass ellipticity, we also allow the
slope of the mass distribution to vary. This is motivated in part by
the recent realization that massive objects have central mass profiles
which depart from the usual scaling solution (Huang \& Broadhurst
1998) becoming flatter in the core than an isothermal model, with a
slope $\gamma <1$ for the projected surface mass density distribution
$\Sigma(\theta)\propto \theta^{-\gamma}$ Fig \ref{last}b shows our
best fit for the lens features.  All are easily reconciled in the
context of this simple model, provided the slope of the mass
distribution is relatively flat ($\gamma\sim 0.4$).  The tolerance on
the slope is roughly $\pm 0.1$ depending on the ellipticity.  These
two parameters can be traded over a limited range and reproduce the
location of the three images and the position angle (PA) of the arc.
Fixing the ellipticity of the arc to match that of the optical light,
$e=b/a=0.7$, requires a slope $\gamma=0.37$.  The central position
and the PA of the mass distribution for the best fit are consistent
with that of the central elliptical galaxy.

  The radial extent or thickness of the arc would be greater for a
flat profile, where the magnification is given by for a power law, so
the fact that object $C$ is resolved supports a flat central slope,
assuming that the source is relatively symmetric. The radio and
optical lensing features can be easily reconciled in more detail when
the structure of an AGN is considered. If the QSO lies just outside
the caustic then only two images of such a small source form in both
the optical and the radio. The outer central region of the host galaxy
must extend over the caustic so that the arc is a third image but only
of the region of the host projected inside the caustic. This must also
be true of the radio emission which lies in the center of the arc;
image $C$ arises naturally in the context of this model from extended
radio emission which originates in a region closer to the QSO than
most of the optical arc light, but is spatially distinct from the QSO
since it must project inside the caustic.

 The best fit model magnification of the radio and optical images of
the QSO images, $A$ and $B$, is 8.5 and 6 respectively.  The magnification
of the radio source $C$ is much greater. Its location in the center of
the arc means that this region of the radio emission straddles the
caustic generating a high overall magnification of $\sim 40$ for the 4
knots. The radio emision could arise from within the generally diffuse
radio emision typically found in the central kpc of an AGN or may be
an image of a jet.  This model is most similar to that of \cite{lan91}
, requiring the source to sit on the caustic, and quite unlike the
multiple component model suggested by \cite{nara}.

  The mass contained within the critical radius is virtually
independent of the mass distribution, $1.6\pm 0.2\times
10^{13}h^{-1}M_{\odot}$, depending somewhat on cosmology through the
ratio of source-lens/source distances, due to the relatively high
redshift of the lens. Taking into account the luminosity of the
central elliptical $L_V^D=4.16\pm 0.21\times 10^{10}h^{-2}L_{\sun}$,
the above value defines a very robust value of the mass--to--light
ratio within the critical radius $M/L_D=
389^{+45}_{-45}h(M/L_V)_{\sun}$,
 
 Interestingly the mass profile advocated by Navarro, Frenk \& White
(1997) tends to a similar flat slope at the critical radius, when
integrated in projection $\Sigma(\theta) \propto \theta^{-0.3}$, and
normalized to the mass contained within the critical radius, for a
reasonable choice of parameters, $r_s=100$ kpc and $d_c=3000$.
Extrapolating to larger radius using this profile yields a mass
$M=1.2^{+0.3}_{-0.3}\times 10^{14}h^{-1}M_{\sun}$ within a radius of
$0.2h^{-1}$ Mpc, assuming circular symmetry, considerably higher than
an extrapolation of a purely isothermal profile, which has a steeper
slope within this radius.  This mass is compatible with the X-ray
estimate of H97, and yields a mass--to--light ratio of $M/L_V=
372^{+94}_{-94}h(M/L_V)_{\sun}$ using our estimate of the cluster
luminosity, where the uncertainty in the mass lies mainly with the
unknown ellipticity of the mass distribution at large radius.


\section{Conclusions}

  We find clear evidence for a distant cluster of galaxies centered on
the QSO lens and luminous X-ray source AXJ2019+112.  A tight red
sequence of galaxies is detected within a radius of $0.2 h ^{-1}$ Mpc
of the known $z=1.01$ elliptical lensing galaxy, with a color and
slope indicative of a cluster at $z=1$. We estimate the restframe
V-band luminosity of the cluster early-types within a $0.4h^{-1}$ Mpc
aperture to be $L_V^E=1.6\pm 0.2\times 10^{11}h^{-2}L_{\sun}$ and the
luminosity from all galaxy types to be $L_V\approx 3.2\times
10^{11}h^{-2}L_{\sun}$. This is an order of magnitude larger
than the luminosity estimations previously published for this
cluster. If we use the X-ray mass estimations of H97 based on the
estimated gas temperature, the mass--to--light ratio within
$0.4 h ^{-1}$ Mpc is $M/L_V=224^{+95}_{-66}h(M/L_V)^{\sun}$, similar
to massive $z\lesssim 0.5$ clusters.  From a weak lensing analysis it
can also be established, independently of the X-ray mass estimate,
that for an isothermal cluster
$$
M/L_V < 520 h \left(\frac{\sigma}{1300 \;
{\rm km \; s^{-1}}}\right)^{2} \left(\frac{\beta}{0.2}\right)^{-1}
\left(\frac{r}{ 200 \; h^{-1}\;{\rm kpc}}\right) h (M/L_V)_{\sun}
$$

  The observation of a radially resolved optical arc in the cluster
center and the detailed model fit to this and the two QSO images
requires that the core of the mass distribution is much flatter than
isothermal. This analysis allows us to set a robust value of the
mass--to--light ratio within the critical radius $M/L_V=
389^{+53}_{-53}h(M/L_V)_{\sun}$. The inferred surface mass
distribution is very flat, $\Sigma\propto \theta^{-0.3}$, and
consistent with the Navarro, Frenk \& White (1997) profile for
reasonable choices of the core radius and central density. An
extrapolation of this profile to larger radius yields a
mass--to--light ratio $M/L_V= 372^{+94}_{-94}h(M/L_V)_{\sun}$ within a
$200 h ^{-1}$ kpc aperture.

  We have reanalyzed the HRI X-ray data and raise the possibility that
the X-ray luminosity of the cluster should be significantly revised
downward. A clear source of emission is spatially coincident with the
cluster center. Two other sources of emission are probably unrelated,
but are smoothed together in the map of Hattori et al (1997)
comprising 50\% of their X-ray flux estimate. Furthermore we cannot
exclude the possibility that the lensed QSO makes a significant
contribution to the X-ray flux in the center. Higher quality spatially
resolved X-ray spectroscopy is needed to understand the relation
between the large beam measurement of a hot thermal spectrum by ASCA
and the clumpy nature of the higher resolution ROSAT-HRI maps. Deeper
HST images would also be of interest in obtaining a better mass
estimate for this cluster from weak lensing. Less than a handful of
massive clusters are known at $z\simeq 1$, none of which have secure
measurements of mass.

\acknowledgments

 NB acknowledges a Basque Government postdoctoral fellowship. TJB and
NB acknowledge a LTSA grant NASA NAG-3280. F.C. is  supported by contracts ARC
94/99-178   ``Action   de Recherche   Concert\'ee  de   la Communaut\'e
Fran\c{c}aise"~ and P\^ole d'Attraction Interuniversitaire P4/05 (SSTC, 
Belgium). We thank Doug Finkbeiner for providing the high resolution 
extinction estimate.

\newpage
\begin{deluxetable}{lrrrrrrrrr}
\tablewidth{0pc}
\tablecaption{Cluster sequence candidates}
\tablehead{
\colhead{$r\arcsec$}&
\colhead{$I$} &
\colhead{$\delta I$} &
\colhead{$V$} &
\colhead{$\delta V$} &
\colhead{$K_S$} &
\colhead{$\delta K_S$} &
\colhead{$V-I$} &
\colhead{$I-K_S$} &
\colhead{$M_V$} 
}
\startdata
00.0& 21.20&  0.024& 24.38& 0.06& 17.12 & 0.02 & 3.17& 4.08 & -21.89\nl
12.0& 23.06&  0.072& 26.09& 0.23& 19.31 & 0.09 & 3.01& 3.75 &  -20.18\nl
12.2& 22.25&  0.033& 25.40& 0.14& 19.25 & 0.06 & 2.93& 3.20 & -20.87\nl
14.3& 22.92&  0.049& 25.92& 0.23& 19.49 & 0.07 & 2.99& 3.43 &  -20.35\nl
24.7& 21.72&  0.024& 24.91& 0.13& 18.17 & 0.03 & 3.17& 3.55 &  -21.36\nl
24.8& 22.31&  0.032& 25.39& 0.16& 19.15 & 0.06 & 3.06& 2.26 & -20.88\nl
25.9& 23.54&  0.082& 26.71& 0.28& 19.85 & 0.11 & 3.14& 3.69 & -19.56\nl
38.4& 23.79&  0.095& 26.61& 0.26& $>21.70$& ---- & 2.81&$<2.09$ & -19.66\nl
39.0& 23.63&  0.084& 26.77& 0.30& 19.85 & 0.09 & 3.12& 3.78 & -19.50\nl
\enddata
\end{deluxetable}

\newpage

\begin{deluxetable}{lrrrr}
\tablewidth{0pc}
\tablecaption{Objects around galaxy D in the I band frame}
\tablehead{
\colhead{Object} & \colhead{$I$} &
\colhead{$\delta I$} &\colhead{$V$} &
\colhead{$\delta V$}
}
\startdata
A &  22.11 & 0.03 & 22.10& 0.01\nl
B &  22.19 & 0.03 & 22.56& 0.01\nl
C &  22.08 & 0.03 & 23.46& 0.02\nl
CA&  22.80 & 0.06 & 25.02& 0.07\nl
D &  21.20 & 0.02 & 24.38& 0.05\nl
\enddata
\end{deluxetable}

\end{document}